\documentclass[11pt]{article}
\usepackage{graphicx,color}
\usepackage{url}
\usepackage{here}
\usepackage{bm}
\usepackage{amsthm}
\usepackage{mathrsfs}
\usepackage{mathtools,stmaryrd}
\usepackage[percent]{overpic}

\usepackage{subcaption}

\usepackage{hyperref}
\usepackage{cite}
\usepackage{latexsym}
\usepackage[english]{babel}
\usepackage{amsmath,amssymb}
\usepackage{MnSymbol}
\usepackage{geometry}
\usepackage{combelow}

\usepackage{todonotes}

\usepackage{tikz}
\usetikzlibrary{shapes.geometric, arrows}

\tikzstyle{block} = [rectangle, draw, text width=8em, text centered, rounded corners, minimum height=3em]
\tikzstyle{line} = [draw, -latex']

\theoremstyle{definition}

\graphicspath{{./figure/}} 

\begin{document}\sloppy

\newcommand{\Figref}[1]{Fig.~\ref{#1}}
\newcommand{\Tabref}[1]{Table~\ref{#1}}
\newcommand{\Bibref}[1]{~\cite{#1}}
% Math definitions
\newcommand{\SO}{\mathrm{SO}}
\newcommand{\so}{\mathfrak{so}}
\newcommand{\bbbr}{\mathbb{R}}
\newcommand{\bbbrn}{\mathbb{R}^n}
\newcommand{\bbbrrr}{\mathbb{R}^3}
\newcommand{\tr}{\mathrm{tr}}
\newcommand{\trace}{\mathrm{trace}}
\renewcommand{\d}{\mathrm{d}}

\newcommand{\const}{\mathrm{const}}
\newcommand{\Fe}{\mathcal{F}_\mathrm{E}}
\newcommand{\Fb}{\mathcal{F}_\mathrm{V}}
\newcommand{\mdef}{:=}
\newcommand{\ex}{\mathrm{e}_x}
\newcommand{\ez}{\mathrm{e}_z}
\newcommand{\ad}{\mathrm{ad}}
\newcommand{\Ad}{\mathrm{Ad}}
\newcommand{\andrm}{~ \mathrm{and} ~}
\newcommand{\grad}{\mathrm{grad}}
\newcommand{\Exp}{\mathrm{Exp}}
\newcommand{\lpartial}{\bar\partial}
\newcommand{\En}{\mathcal{E}}
\newcommand{\g}{\mathfrak{g}}
\newcommand{\gad}{\mathfrak{g}^*}

\newcommand{\inverse}{^{-1}}
\newcommand{\trans}{^{\top}}
\newcommand{\skewmap}{^{\times}}
\newcommand{\unskewmap}{^{\vee}}
\newcommand{\norm}[1]{\left\lVert#1\right\rVert}
\newcommand{\normtwo}[1]{\left\lVert#1\right\rVert_2}
\newcommand{\normf}[1]{\left\lVert#1\right\rVert_F}
\newcommand{\normx}[1]{\left\lVert#1\right\rVert_x}
\newcommand{\normxk}[1]{\left\lVert#1\right\rVert_{x_k}}
\newcommand{\inpro}[1]{\left\langle#1\right\rangle}

%Editing def
\newcommand{\mathcolorbox}[2]{\colorbox{#1}{$\displaystyle #2$}}

\title{Analysis of molecular dynamics simulation data via statistical distances between covariance matrices}

\author{Yusuke Ono$^{1}$\thanks{Email: yuu555yuu@keio.jp}, \ 
Takumi Sato$^{2}$\thanks{Email: sato8322@keio.jp}, \ 
Kenji Yasuoka$^{2}$\thanks{Email: yasuoka@mech.keio.ac.jp}, \  and \ 
Linyu Peng$^{2}$\thanks{Corresponding author. Email: l.peng@mech.keio.ac.jp} 
\vspace{0.4cm}
\\
{\it 1. Graduate School of Science and Technology, Keio University,}\\
{\it Yokohama 223-8522, Japan}\\
{\it 2. Department of Mechanical Engineering, Keio University,} \\
{\it Yokohama 223-8522, Japan}\\
}

\maketitle

\begin{abstract}
    Molecular dynamics (MD) simulations are powerful tools for elucidating the macroscopic physical properties of materials from microscopic atomic behaviors. 
    However, the massive, high-dimensional datasets generated by MD simulations pose a significant challenge for analysis, necessitating efficient dimensionality reduction and feature extraction techniques.
    While existing methods such as principal component analysis and unsupervised learning have been utilized, issues regarding data efficiency and computational cost remain. 
    In this study, we propose a statistical analysis framework focusing on the analysis of the particle data distributions through their covariance matrices, 
    corresponding to the second-order moments of MD trajectory data. Discrepancies between system states are quantified using statistical distances between these covariance matrices. By applying dimensionality reduction to the resulting distance matrix, we extract lower-dimensional features that characterize the systems' dynamics. 
    We validate the proposed method using Lennard-Jones (LJ) particle systems under different temperature conditions, as well as 
    separate bulk systems of ice and liquid water. 
    The results of LJ particles demonstrate an approximately linear correlation between the first principal component obtained through dimensionality reduction of the distance matrix and the diffusion coefficient. 
    This suggests that global physical properties can be effectively inferred from local statistical information, such as covariance matrices, offering a data-efficient alternative for analyzing complex molecular systems.
    Furthermore, in the case of separate bulk systems of ice and liquid water,
    the method successfully distinguishes between the two phases, highlighting its potential for characterizing phase transitions and structural differences in molecular systems.
\end{abstract}

\section{Introduction}

Molecular dynamics (MD) simulations have established themselves as indispensable tools in modern materials science and physics. 
By numerically integrating the equations of motion of atoms and molecules, 
MD simulations enable the investigation of material properties at atomic resolution, thereby bridging the gap between microscopic interactions and macroscopic observables \cite{karplus2002molecular, dror2012biomolecular, allen2017computer}. 
This approach has been successfully applied to a wide range of systems, from simple fluids to complex biomolecules, 
allowing the evaluation of mechanical, thermal, electrical, and magnetic properties that are often difficult to access experimentally \cite{bock2023simulation, sponer2018rna}.
However, advances in high-performance computing have led to a rapid growth in the volume of MD data. 
Modern simulations routinely generate massive datasets containing the positions and velocities of thousands to millions of particles over long time scales. 
Extracting meaningful and interpretable physical insights from such high-dimensional data remains computationally demanding and nontrivial. 
Consequently, there is a pressing need for efficient data analysis frameworks that can reduce the dimensionality of MD trajectories while preserving the essential dynamics governing macroscopic behavior.

To address the challenges posed by high-dimensional data, various dimensionality reduction and feature extraction techniques have been introduced in computational physics. These techniques are generally categorized as linear or nonlinear methods \cite{roweis2000nonlinear, cunningham2015linear, sugiyama2016}.
Classical linear approaches such as principal component analysis (PCA) have been widely used to identify collective motions in proteins and to characterize complex energy landscapes \cite{amadei1993ED, kitao1998energy, maisuradze2009pca, david2014pca, jkitao2022pca}.
More recently, advanced signal processing techniques, such as singular spectrum transformation, have shown promise in detecting changes in protein motion modes  \cite{sst4protein2021}.
In contrast, nonlinear methods, including t-distributed stochastic neighbor embedding (t-SNE) \cite{maaten2008visualizing},
uniform manifold approximation and projection \cite{mcinnes2018umap},
and modern unsupervised machine learning algorithms using variational autoencoders \cite{Kingma2014} and generative adversarial networks \cite{gan2014}, 
have been employed to capture intricate, nonlinear relationships in MD data \cite{hradiska2024tsne, trozzi2021umap, Wei2018vaeMD, Endo_Tomobe_Yasuoka_2018, Endo_Yuhara_Tomobe_2019, yasuda2022}.
A key advantage of nonlinear approaches is their ability to preserve more complex structural information than linear methods when reduced to the same dimensionality.

Despite these developments, existing methods often face limitations in terms of data efficiency and interpretability. Many data-driven approaches operate directly on raw trajectory data, which can be computationally expensive and may obscure the underlying physical mechanisms. Moreover, there is often a gap between the geometric features extracted by such algorithms and the thermodynamic or transport properties of interest. Developing a method that is both computationally efficient and physically interpretable therefore remains an open challenge.

In this study, we propose an alternative statistical framework that focuses on covariance matrices derived from MD data, such as particle positions or velocities, which correspond to the second-order moments of the underlying probability distributions.
In statistical mechanics, 
fluctuations---particularly velocity fluctuations---and their correlations are intrinsically related to thermodynamic state variables and transport coefficients \cite{kubo1966fluctuation, mcquarrie2000statistical}. 
Therefore, the covariance matrix provides a compact yet information-rich descriptor for capturing linear correlations within the system \cite{hansen2013theory}.
We acknowledge, as discussed in \cite{lange2006generalized}, that complex MD systems such as biomolecules also exhibit nonlinear correlations. 
Nevertheless, since many thermodynamic quantities are directly linked to second-order moments of velocity, we adopt the covariance matrix as the primary descriptor in this work.

Our methodology centers on quantifying the statistical distance between covariance matrices derived from different time windows or system states.
By constructing a distance matrix that captures the dissimilarity between these states and applying dimensionality reduction to this distance matrix, 
we obtain a lower-dimensional representation of the system's evolution. This approach enables the identification of subtle differences in thermodynamic states that may not be apparent from conventional coordinate-based metrics.
The main contributions of this study are summarized as follows:
\begin{itemize}
    \item We introduce a novel framework for analyzing MD simulation data based on statistical distances between covariance matrices of particle time-series data.
    \item We demonstrate that this approach effectively captures the system's dynamics and enables dimensionality reduction while preserving essential physical information.
    \item We validate the proposed method using two types of MD datasets. First, we apply it to a Lennard-Jones (LJ) particle system and demonstrate a clear correlation between the extracted lower-dimensional features and macroscopic transport properties, such as the diffusion coefficient. Second, we investigate bulk systems of ice and liquid water, showing that the method effectively distinguishes between different molecular phases.
\end{itemize}

The remainder of this paper is organized as follows. In Section \ref{sec: methodology}, we present the theoretical foundation of the statistical distance framework 
and describe the computational procedures and algorithms employed.
In Section \ref{sec: results}, we apply the proposed method to LJ particle systems at various temperatures, as well as to separate bulk systems of ice and liquid water, and discuss the relationship between the extracted features and physical properties.
Finally,  Section \ref{sec: conclusions} summarizes our findings and outlines potential directions for future research.
  
  \section{Methods}
\label{sec: methodology}

In this section, we introduce the proposed method for analyzing MD data using statistical distances between estimated covariance matrices. 
To establish its statistical foundation, we begin with the method of moments, a classical estimation framework 
in which the parameters of a probability distribution are obtained by equating sample moments with their theoretical counterparts \cite{Pearson1893}.

Let $\bm{W}\in\mathbb{R}^d$ be a random vector with some probability density function $f(\bm{W})$. The $n$-th moment of the distribution is defined as
\begin{equation}
    \mathbb{E}[\bm{W}^{\otimes n}],
\end{equation}
provided that the expectation exists, where $\mathbb{E}[\cdot]$ is the statistical expectation and $\otimes$ denotes the tensor product. In particular, the first-order moment corresponds to the mean vector,
\begin{equation}
    \mathbb{E}[\bm{W}],
\end{equation}
while the second-order central moment corresponds to the covariance matrix
\begin{equation}
    \mathbb{E}[(\bm{W}-\mathbb{E}[\bm{W}])(\bm{W}-\mathbb{E}[\bm{W}])^\top].
\end{equation}

Now let $L$ observations $\{\bm{x}_1, \bm{x}_2, \ldots, \bm{x}_{L}\}$ be given,  where each $\bm{x}_k \in \mathbb{R}^d$. The first-order and second-order moments, i.e., the mean vector and the covariance matrix, can be estimated from the observations as
\begin{equation}
    % \overline{\bm{x}} = \frac{1}{n} \sum_{i=1}^{n} \bm{x}_i ,
    \bm{\mu} = \frac{1}{L} \sum_{i=1}^{L} \bm{x}_i ,
\end{equation}
and
\begin{equation}
    % \bm{\Sigma} = \frac{1}{n} \sum_{i=1}^{n} (\bm{x}_i - \overline{\bm{x}})(\bm{x}_i - \overline{\bm{x}})^\top .
    \bm{\Sigma} = \frac{1}{L} \sum_{i=1}^{L} (\bm{x}_i - \bm{\mu})(\bm{x}_i - \bm{\mu})^\top.
\end{equation}
Under the assumption of multivariate normality, this sample covariance matrix (SCM) coincides with the maximum likelihood estimator of the covariance matrix \cite{SCMgoodman}.
A known limitation of the SCM is that its accuracy critically depends on the availability of a sufficiently large number of independent and identically distributed (i.i.d.) observations \cite{reed1974}. 
In our framework, the estimators of the mean vector and covariance matrix serve as fundamental statistical descriptors for the multidimensional analysis of MD trajectories.

We now describe how covariance matrices are constructed from time-series data obtained through MD simulations.
Let 
\begin{equation}
    \left\{\bm{x}_k=(x_k^1,x_k^2,x_k^3)\in\mathbb{R}^3\right\}
\end{equation} 
be a series of data (i.e., positions or velocities) obtained from simulations or experiments; $k$ means discrete time physically, 
i.e., $\bm{x}_k=\bm{x}(k \Delta t)$ where $\Delta t$ is the timestep. 
The series is divided into small pieces, referred to as sub-windows, each of length $N$, as follows
\begin{equation}
    \{(\bm{x}_1,\bm{x}_2,\ldots,\bm{x}_N), \ (\bm{x}_{N+1},\bm{x}_{N+2},\ldots,\bm{x}_{2N}),\ \ldots  \} .
\end{equation}
Each small piece can be written as a $3\times N$ matrix (with vectors in $\mathbb{R}^3$ expressed as columns), and we assume that there are $K$ numbers of them, namely
\begin{equation*}\label{eq:matdata}
    \bm{X}_1=[\bm{x}_1,\bm{x}_2,\ldots,\bm{x}_N],\quad 
    \bm{X}_2=[\bm{x}_{N+1},\bm{x}_{N+2},\ldots,\bm{x}_{2N}], 
    \ \ldots, \ 
    \bm{X}_K=[\bm{x}_{(K-1)N+1},\bm{x}_{(K-1)N+2},\ldots,\bm{x}_{KN}].
    % \bm{X}_1=(\bm{x}_1,\bm{x}_2,\ldots,\bm{x}_N),\quad 
    % \bm{X}_2=(\bm{x}_{N+1},\bm{x}_{N+2},\ldots,\bm{x}_{2N}), 
    % \quad \ldots, \quad 
    % \bm{X}_K=(\bm{x}_{(K-1)N+1},\bm{x}_{(K-1)N+2},\ldots,\bm{x}_{KN}).
\end{equation*}
Therefore, $\bm{X}_m \in \mathbb{R}^{3 \times N}$ denotes the data matrix for the $m$-th segment ($1 \le m \le K$). For each segment $\bm{X}_m$, we omit the index $m$ when no confusion arises and denote its components as
\begin{equation}
    \begin{pmatrix}
        x_{1}^1 & x_{2}^1 & \cdots & x_{j}^1 & \cdots & x_{N}^1 \vspace{0.15cm} \\
        x_{1}^2 & x_{2}^2 & \cdots & x_{j}^2 & \cdots & x_{N}^2 \vspace{0.15cm}\\
        x_{1}^3 & x_{2}^3 & \cdots & x_{j}^3 & \cdots & x_{N}^3
    \end{pmatrix},
\end{equation}
where $x_{j}^\alpha$ represents the value of the $\alpha$-th spatial component with $\alpha=1, 2, 3$ corresponding to $x, y, z$, and 
at the $j$-th time step within the $m$-th segment ($1 \le j \le N$).
Specifically, the $j$-th column corresponds to the data vector $\bm{x}_{(m-1)N + j}$.
The corresponding covariance matrix $\bm{R}_m \in \mathbb{R}^{3N \times 3N}$ for the $m$-th segment, $\bm{X}_m $, is assembled by arranging these blocks:
\begin{equation}\label{Rm}
    \bm{R}_m = 
    \begin{pmatrix} 
        R_{xx} & R_{xy} & R_{xz} \\
        R_{yx} & R_{yy} & R_{yz} \\
        R_{zx} & R_{zy} & R_{zz} 
    \end{pmatrix}.
\end{equation}
For each block $R_{\alpha\beta}$, we impose a Toeplitz structure. Although time-series data in practical applications is not necessarily stationary, 
it has been widely demonstrated in signal processing that enforcing a Toeplitz structure enhances the accuracy and robustness of covariance matrix estimation \cite{ABY2012,Du2020,Zatman2001}.
Motivated by these findings, we construct the $N \times N$ block matrix $R_{\alpha\beta}$ in Toeplitz form as follows \cite{Arnaudon2013, Cabanes2019}:
\begin{equation}
    R_{\alpha\beta} = 
    \begin{pmatrix}
        r_0^{\alpha\beta} & \cdots & r_k^{\alpha\beta} & \cdots & r_{N-1}^{\alpha\beta} \\
        \vdots & \ddots & \ddots & \ddots &  \vdots \\
        r_k^{\beta\alpha} & \ddots & r_0^{\alpha\beta} & \ddots & r_{k}^{\alpha\beta} \\
        \vdots & \ddots & \ddots & \ddots & \vdots \\
        r_{N-1}^{\beta\alpha} & \cdots & r_k^{\beta\alpha} & \cdots & r_0^{\alpha\beta}
    \end{pmatrix}.
    \label{eq:block_toeplitz}
\end{equation}
This definition implies that $R_{\alpha\beta} = R_{\beta\alpha}\trans$, ensuring that $\bm{R}_m$ defined in \eqref{Rm} is symmetric.
The correlation function $r_k^{\alpha\beta}$ between spatial components 
$\alpha$ and $\beta$ ($\alpha,\ \beta =1, 2, 3$) at lag $k$ ($0 \le k \le N-1 $) is given by 
\begin{equation}\label{eq:correlation_function}
		r_k^{\alpha\beta} = \mathbb{E} \left[x_{l}^\alpha\ x_{l+k}^\beta \right], \text{ for any } 0 \le l \le N-k-1,
\end{equation}
which can be estimated by the observation data as
\begin{equation}\label{eq: estimator}
    r_k^{\alpha\beta} =  \frac{1}{N} \sum_{l=0}^{N-k-1} x_{ l}^\alpha \, x_{l+k}^\beta , \quad 0 \le k \le N-1.
\end{equation}
Note that $r_0^{\alpha\beta}= r_0^{\beta\alpha}$ for all $\alpha,\beta$. 
This then yields an estimate for each $R_{\alpha\beta}$ in \eqref{eq:block_toeplitz} and, hence, $\bm{R}_m$ in \eqref{Rm}.
Under standard conditions (e.g., non-degenerate stochastic excitation and sufficiently rich data), it is positive definite and therefore full rank with probability one.
The set of symmetric positive definite (SPD) matrices forms a Riemannian manifold, known as the SPD manifold (e.g., \cite{bhatia2009positive, bhatia2019165}).
In this study, for simplicity, we adopt the Euclidean distance as the statistical distance between two SPD covariance matrices $\bm{R}_i, \bm{R_j}$:
\begin{equation}\label{dis}
    d\left(\bm{R}_i, \bm{R}_j\right) = \left\| \bm{R}_i - \bm{R}_j \right\|_F,
\end{equation}
where $\| \bm{R} \|_F = \sqrt{\trace \left(\bm{R} \bm{R}\trans \right)}$ is the Frobenius norm.
Accordingly, the Euclidean mean of the set of SPD matrices $\{\bm{R}_1,\bm{R}_2,\ldots,\bm{R}_K\}$ is defined as their arithmetic mean
\begin{equation}
\frac{\bm{R}_1+\bm{R}_2+\cdots +\bm{R}_K}{K}.
\end{equation}

\begin{figure}[htbp]
    \centering
\includegraphics[width=0.85\linewidth]{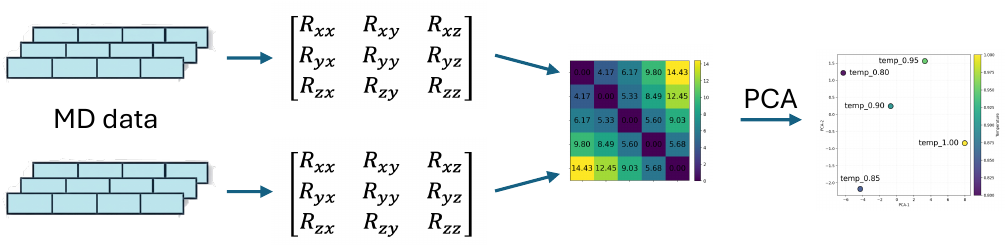}
    \caption{Schematic illustration of the proposed method.}
    \label{fig:Schematic_figure}
\end{figure}

\begin{figure}[htbp]
    \centering
    \includegraphics[width=.5\linewidth]{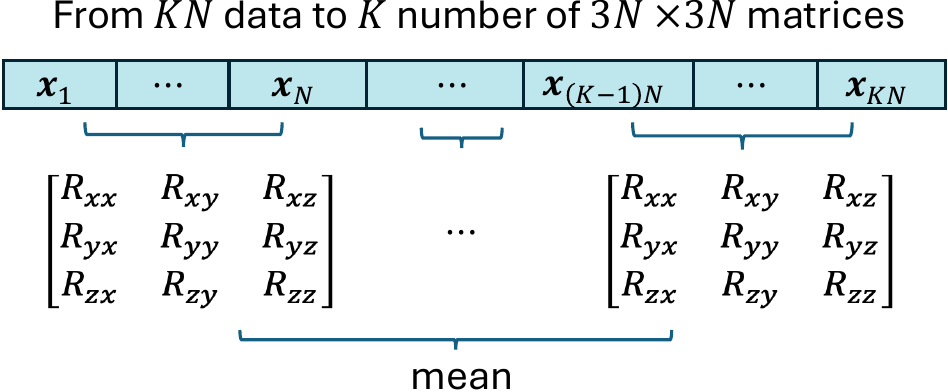}
    \caption{Computation for the covariance matrices.}
    \label{fig:Computation_covariance_matrix}
\end{figure}

To analyze the MD data, we characterize the state of the system by computing statistical distances between these matrices.
The overall procedure of the proposed method is illustrated in \Figref{fig:Schematic_figure}.
First, the raw time-series data are normalized, if necessary, to ensure numerical consistency. 
The entire dataset is then partitioned into $K$ segments of length $N$, and for each segment, a $3N \times 3N$ block covariance matrix is constructed. 
These matrices consist of nine $N \times N$ blocks that capture the temporal covariances between spatial components, as defined in \eqref{eq:block_toeplitz}.
To obtain a robust statistical representation,
the ensemble (Euclidean) mean of these covariance matrices is computed over all $K$ segments as shown in \Figref{fig:Computation_covariance_matrix}.
Subsequently, the dissimilarities between different datasets are quantified by calculating the Euclidean distances between their respective mean matrices. 
Dimensionality reduction is then performed using PCA, projecting the data onto a two-dimensional space while preserving the resulting distance matrix.
The embeddings of the systems are thereby obtained.
The lower-dimensional geometry of these embeddings captures the intrinsic variables driving the variations between systems.
Finally, the physical significance of the embedding space is investigated by correlating the principal components with the physical properties of the system.

\section{Results and Discussions}
\label{sec: results}

In this section, we apply the proposed method to two systems: an LJ particle system and separate bulk systems of ice and liquid water, respectively. %, to present the numerical results obtained by our proposed method.

\subsection{Lennard-Jones systems}
The LJ potential is one of the most widely used models for describing interactions in monatomic fluids. 
In this study, we validate the proposed method by analyzing time-series data obtained from MD simulations of a particle system governed by the LJ potential. 
Specifically, we consider the LJ $12-6$ potential characterized by the length scale $\sigma$ and the energy scale $\varepsilon$, 
\begin{equation}
U(r) = 4\varepsilon \left( \left( \frac{\sigma}{r} \right)^{12}
- \left( \frac{\sigma}{r} \right)^{6} \right).
\end{equation}

All quantities are expressed in reduced LJ units with $\sigma = 1$, $\varepsilon = 1$, and particle mass $m = 1$.
The system consists of $4,\!000$ particles, and the initial configuration is generated using a face-centered cubic (FCC) lattice. 
An equilibrated configuration is obtained by performing a preliminary simulation at temperature $T = 1.0$ using a Langevin thermostat, during which the simulation box is allowed to relax. 
After equilibration, the box length converges to $17.1$, corresponding to a number density of $\rho \simeq 0.8$. 
This equilibrated configuration is used in the subsequent production runs.
The time evolution of the particles is computed using the velocity Verlet algorithm, with periodic boundary conditions imposed in all three spatial directions. Simulations are performed at five nondimensional temperatures, $T = 0.80,\ 0.85,\ 0.90,\ 0.95$, and $1.00$.
For each temperature, velocity time series are collected in the microcanonical (NVE) ensemble and used as input data for the statistical distance analysis proposed in this study. The time step is set to $\Delta t = 0.005$, and trajectory data over $100{,}000$ time steps are used for the analysis.

\begin{figure}[htbp]
    \centering
    \includegraphics[width=0.45\linewidth]{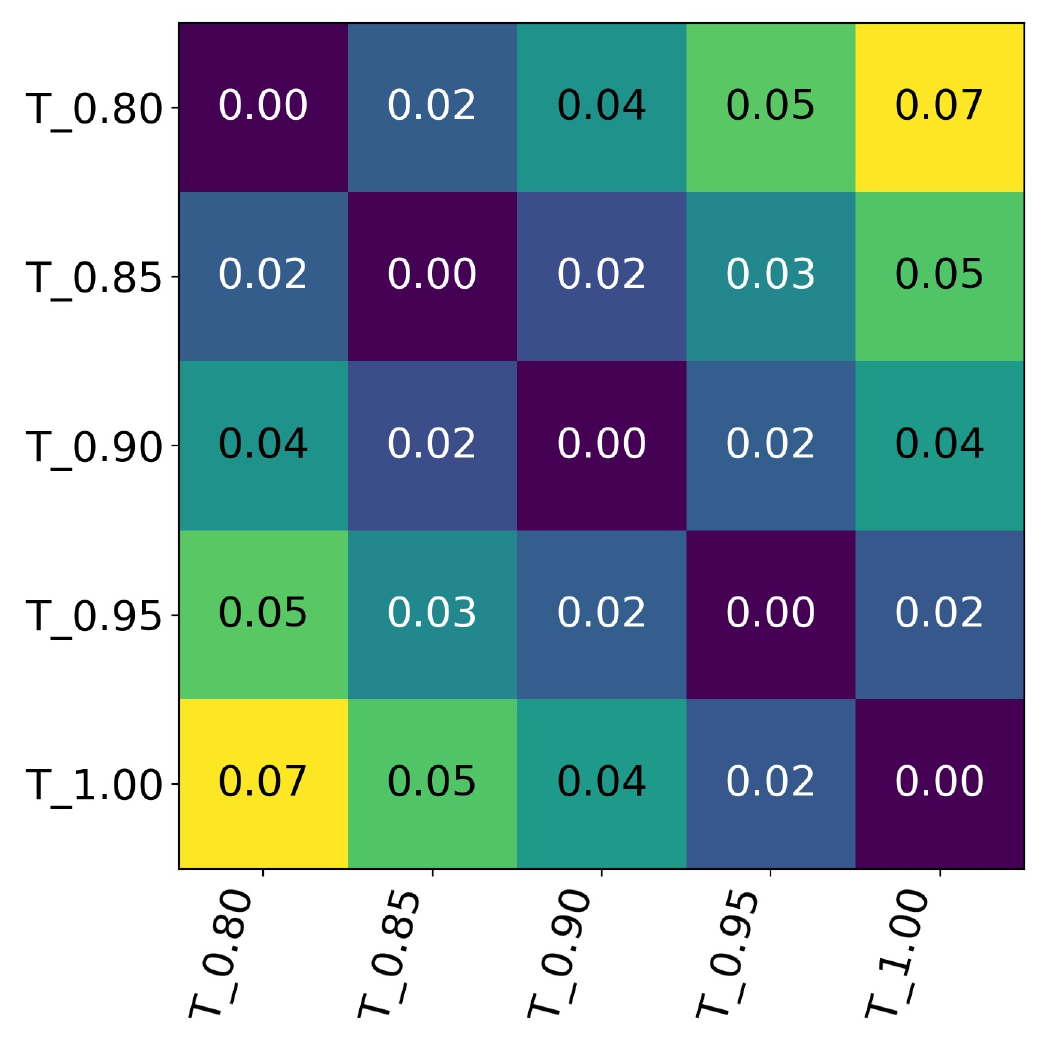}
    \caption{Distance matrix between covariance matrices at different temperatures.}
    \label{fig:Distance_matrix}
\end{figure}

In this experiment, we focus on analyzing the velocity data of particles obtained from the MD simulations 
because of the stationarity of the velocity time series under the theoretical framework considered.
Before evaluating the distances between the covariance matrices, the velocity data are normalized within each small segment of $N$.
The distance matrix evaluated with $N=8$ and $K=12,\!500$ is shown in \Figref{fig:Distance_matrix}. It can be observed that the distances increase as the differences between the states become larger.

\begin{figure}[htbp]
    \centering
    \includegraphics[width=0.45\linewidth]{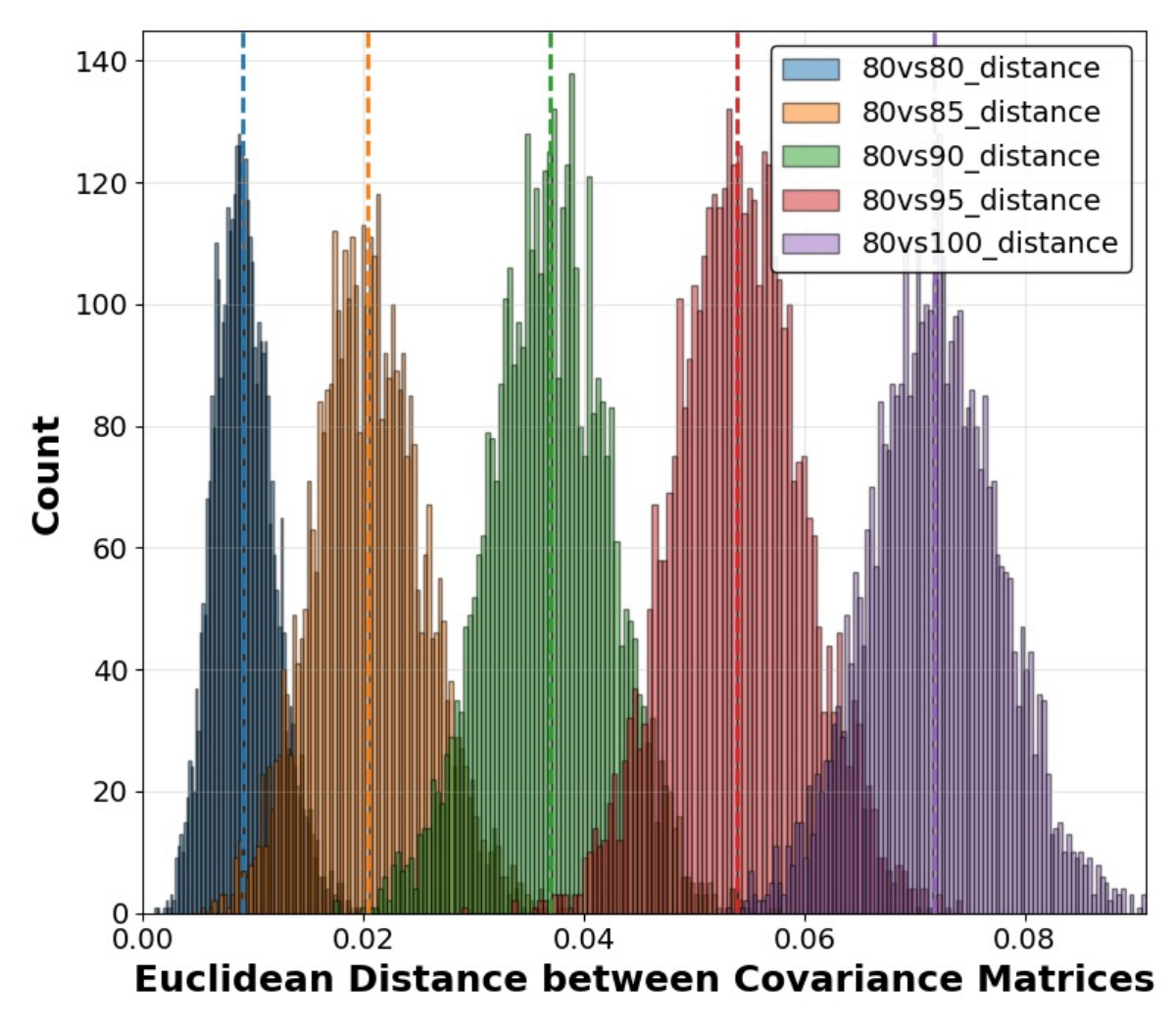}
    \caption{The histograms of the distances between $T=0.80$ and the other temperatures. Each color corresponds to a specific temperature: $T=0.80$ (blue), $0.85$ (orange), $0.90$ (green), $0.95$ (red), and $1.00$ (purple).}
    \label{fig: hist_lj}
\end{figure}

\begin{figure}[htbp]
    \centering
    \includegraphics[width=0.45\linewidth]{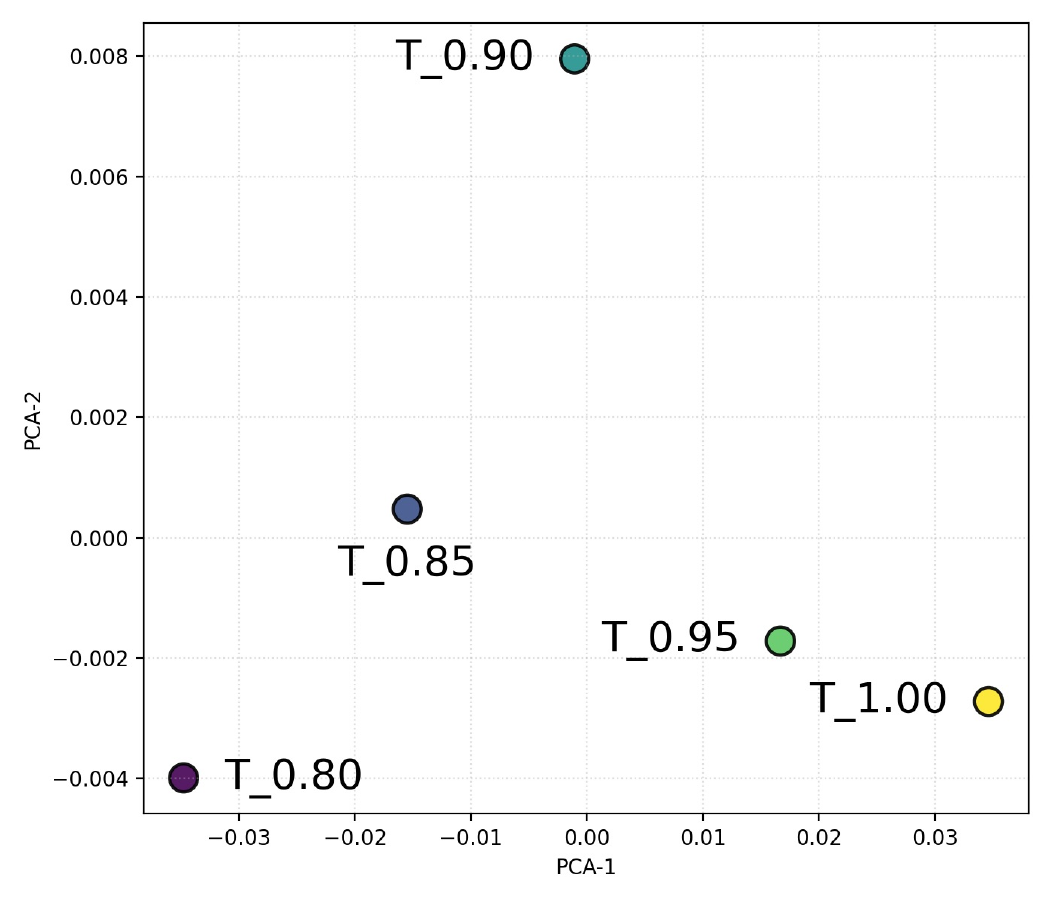}
    \caption{The two-dimensional PCA projection of the distance matrix.}
    \label{fig:velocity_pca2}
\end{figure}

While this matrix represents the results for a single specific pair of two different molecules, 
a more statistically robust analysis was conducted by randomly selecting $4,\!000$ pairs 
for each temperature, including $T=0.80$ itself, with comparisons made relative to $T=0.80$.
The resulting histograms are shown in \Figref{fig: hist_lj} and exhibit distinct distributions, indicating that temperature variations can be detected through these statistical distances.
By embedding the distance matrix into a two-dimensional space by PCA,  
the result presented in \Figref{fig:velocity_pca2} shows that the data points are ordered according to temperature along the first principal component (PC1) axis.
This demonstrates that the proposed method successfully captures the key molecular behavior underlying the temperature differences.

\begin{figure}[htbp]
    \centering
    \includegraphics[width=0.45\linewidth]{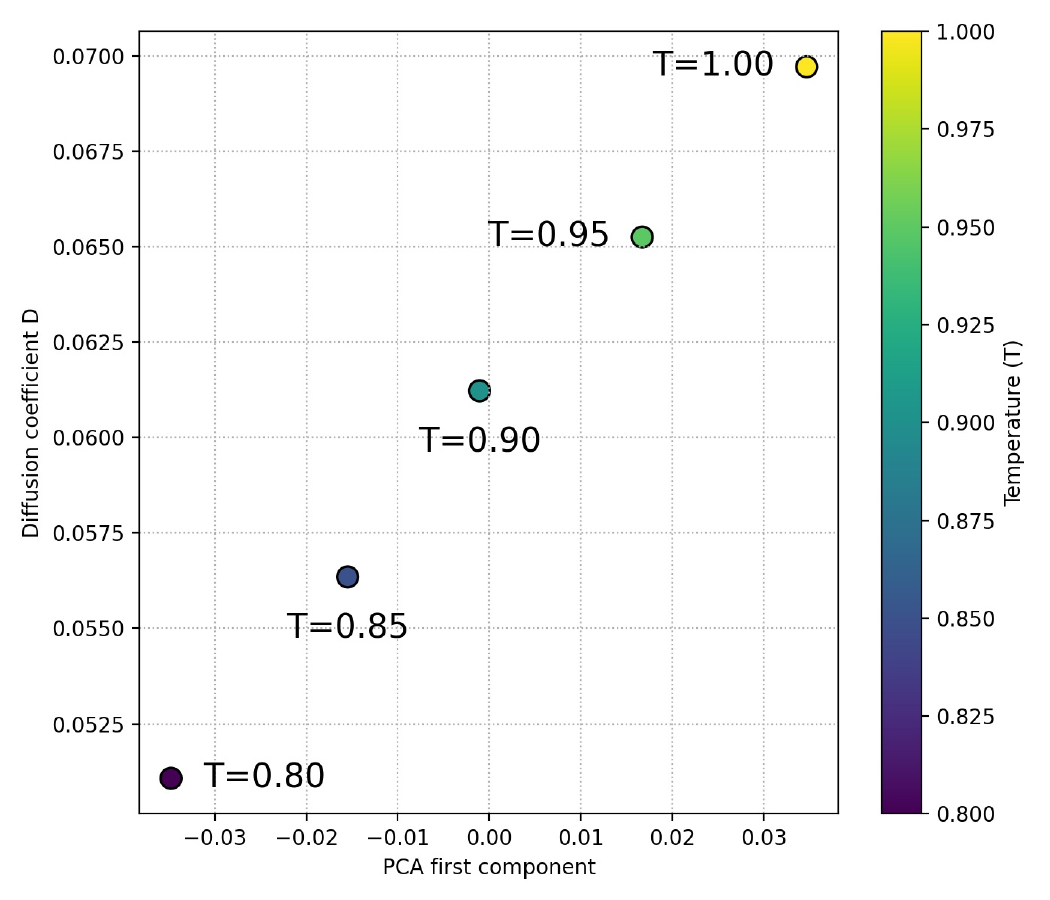}
    \caption{The relationship between the first principal component and the diffusion coefficient.}
    \label{fig:velocity_pc1_vs_diffusion}
\end{figure}

To further quantify this relationship, we examine the correlation between PC1 and the diffusion coefficient calculated from the velocity data.
As shown in \Figref{fig:velocity_pc1_vs_diffusion}, a clear linear relationship is observed between PC1 and the diffusion coefficient. These results indicate that the diffusion coefficient of LJ particle systems across different temperatures can be effectively estimated from local statistical information derived from only eight consecutive time steps.

\subsection{Separate bulk systems of ice and liquid water}

\begin{figure}[htbp]
\centering
\begin{subfigure}{0.45\linewidth}
    \centering
    \includegraphics[width=0.60\linewidth]{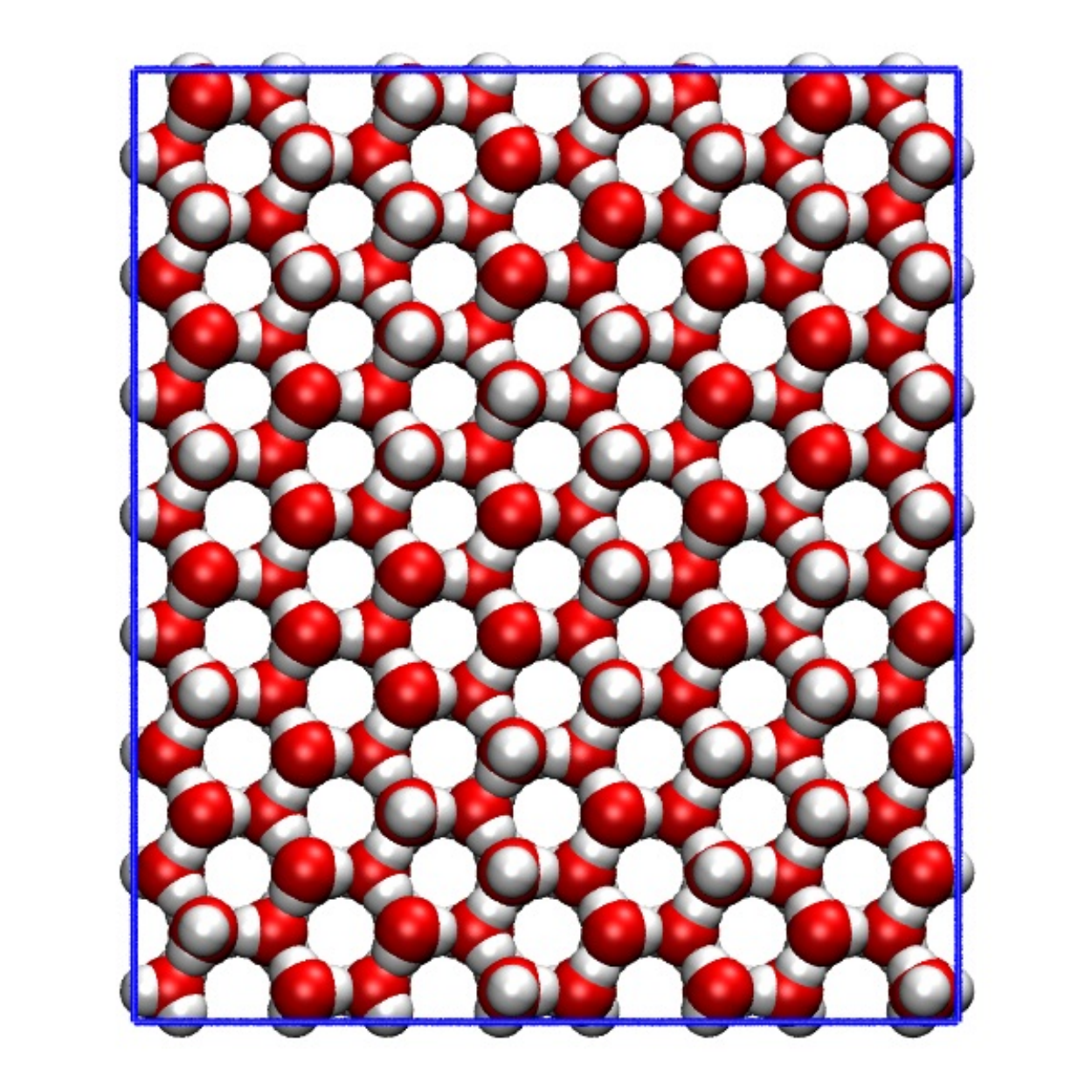}
    \caption{Bulk ice}
\end{subfigure}
\hspace{0.02\linewidth}
\begin{subfigure}{0.45\linewidth}
    \centering
    \includegraphics[width=0.60\linewidth]{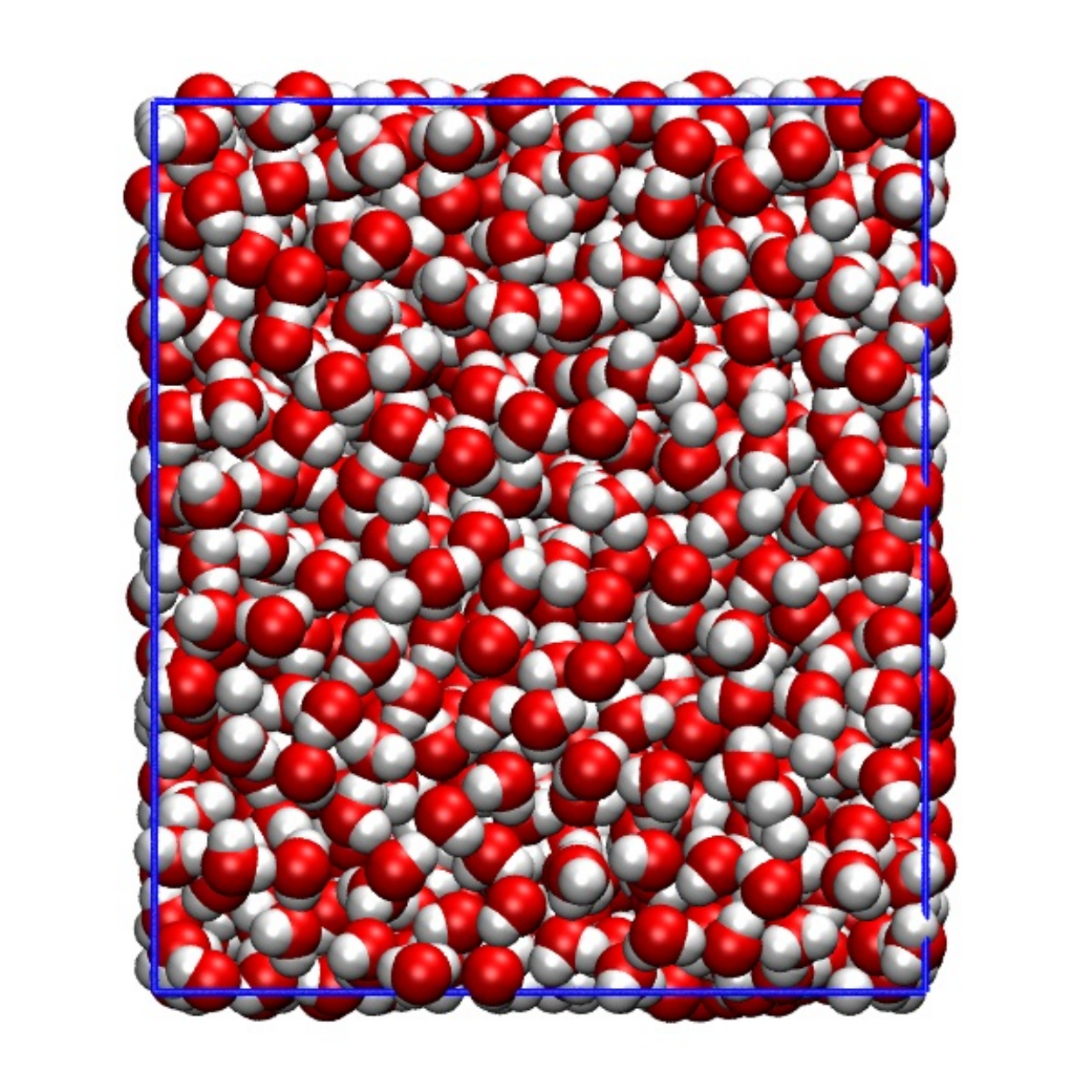}
    \caption{Bulk liquid water}
\end{subfigure}
\caption{Each initial configurations of the systems consists of $1,\!024$ TIP4P/Ice water molecules.}
\label{fig:gro_icewater}
\end{figure}

In the second application, we analyze dipole moment vectors obtained from MD simulations of two separate systems: bulk ice and bulk liquid water. 
The intermolecular interactions between water molecules are described using the TIP4P/Ice water model \cite{tip4pice2005},
a rigid and non-polarizable water model that is known to reproduce the melting point and phase behavior of ice. 
Each system consists of $N = 1,\!024$ water molecules placed in a cubic simulation box with periodic boundary conditions imposed in all three spatial directions. 
All simulations were performed using \textsc{GROMACS 2024.6} \cite{Abraham2015}.
The initial configuration of the ice system was generated using GenIce \cite{genice}, which produces a proton-disordered ice Ih structure, as shown in \Figref{fig:gro_icewater}(a). 
In contrast, the liquid water system is prepared by randomly placing water molecules in the simulation box, as shown in \Figref{fig:gro_icewater}(b). 
The systems were first equilibrated in the NPT ensemble for $10\,\mathrm{ns}$ at $T = 269.0\,\mathrm{K}$ and $P = 0.1\,\mathrm{MPa}$ 
using the Nos\'e--Hoover thermostat \cite{Nose1984JCP,Hoover1985PRA} and the Berendsen barostat \cite{Berendsen1984}.
Production simulations were then performed in the NVT ensemble for $5\,\mathrm{ns}$. 
The equations of motion were integrated using the velocity Verlet algorithm with a time step of $\Delta t = 1\,\mathrm{fs}$. 
Dipole moment vectors were recorded every $10\,\mathrm{fs}$ for subsequent analysis.

\begin{figure}[htbp]
    \centering
    \includegraphics[width=0.75\linewidth]{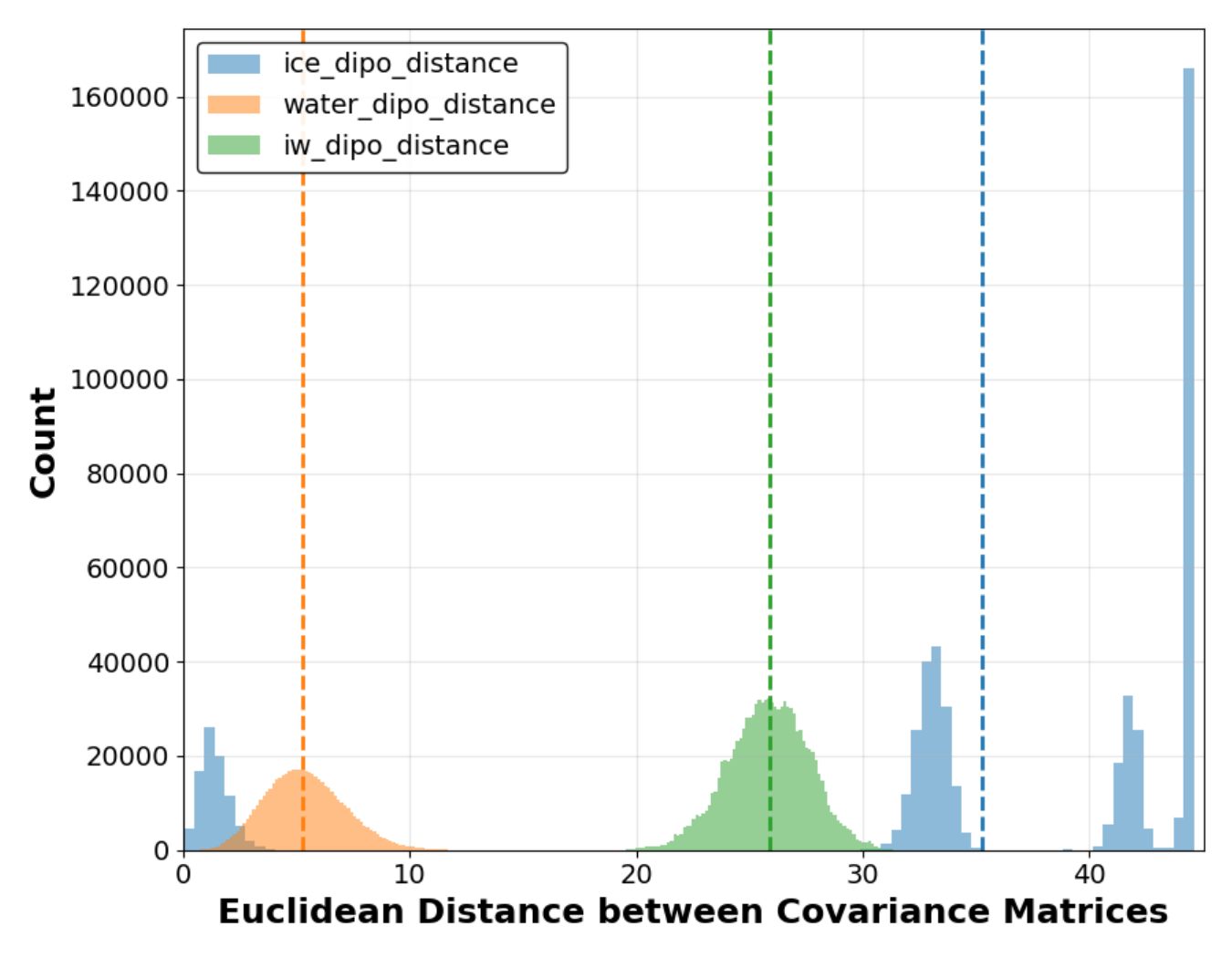}
    \caption{The three histograms show the distances between the ice system and the liquid water system: distances between ice molecules (blue), distances between liquid water molecules (orange), and distances between the two systems (green). The dotted lines in each color indicate the mean values of the corresponding histograms.}
    \label{fig:hist_ice_water_wo_norm}
\end{figure}

Here, we conduct the experiment both with and without data normalization.
In either case, to ensure a statistically robust comparison, we analyze histograms of the distances between covariance matrices. 
Specifically, these histograms are generated by calculating the distances of all pairs for all combinations of states.
The histograms of the distances between covariance matrices for the ice and liquid water systems are evaluated with $N=8$ and $K=12,\!500$ for both cases. 
For raw data without normalization, the histograms of the distances, shown in \Figref{fig:hist_ice_water_wo_norm}, indicate that the two states can be distinguished by the distances between covariance matrices
based on the distinct shapes and positions of these distance distributions when viewed from the perspectives of both liquid water and ice molecules. 
This distinction may be attributed to the characteristic timescales of dipole moment oscillations. In the ice phase, these oscillations occur at higher frequencies.
Consequently, some pairs of ice molecules exhibit similar correlation patterns, while others do not, resulting in a broader distribution of distances even within the same phase. 
However, liquid water molecules exhibit lower-frequency oscillations that produce more uniform and consistent correlation patterns. 
As a result, the histograms are concentrated at smaller distance values, effectively blurring the boundary between the two states when water is used as the reference.

\begin{figure}[htbp]
    \centering
    \includegraphics[width=0.5\linewidth]{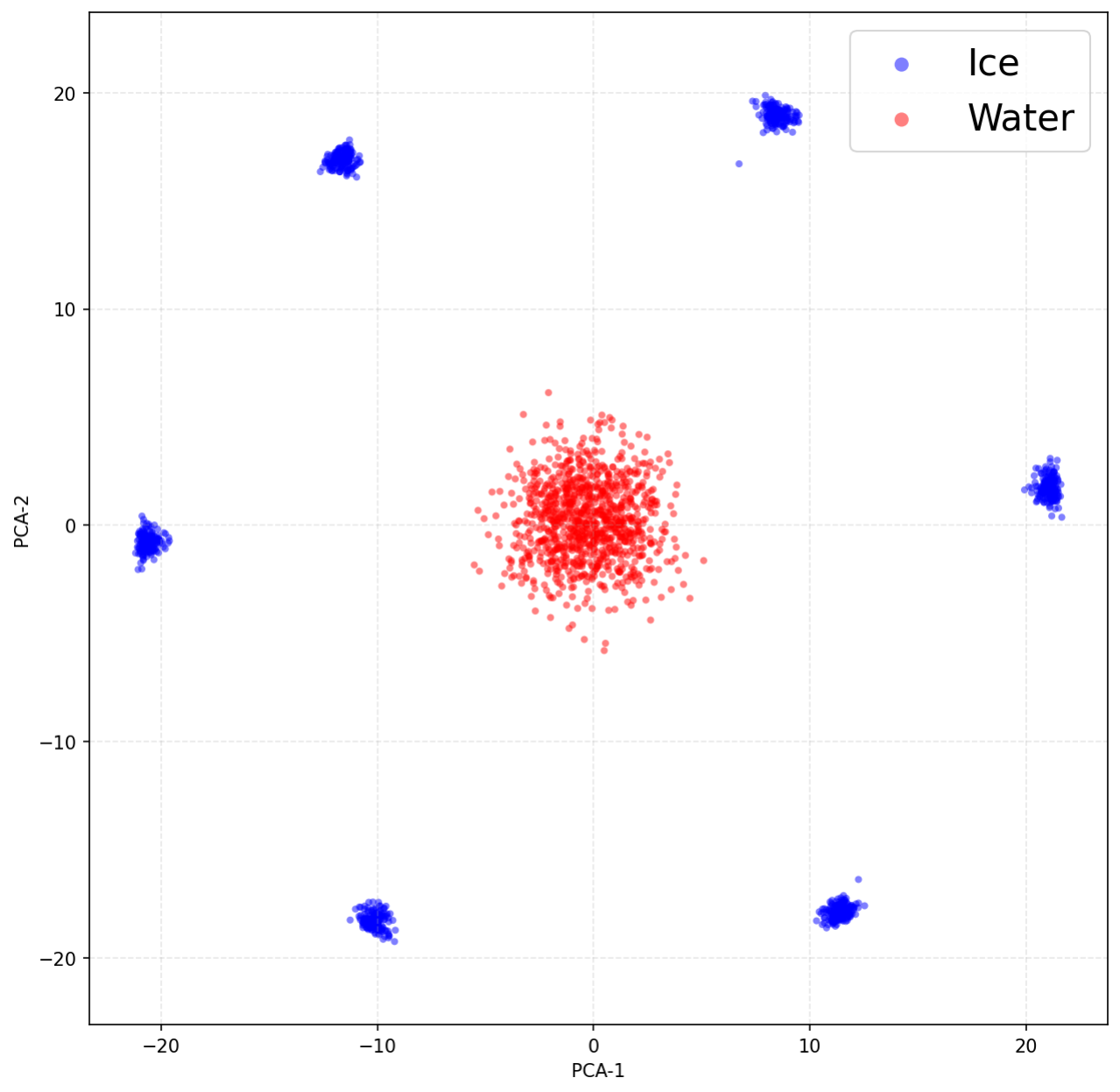}
    \caption{The two-dimensional PCA projection of all pairwise distances within and between water (red) and ice (blue) molecules without  data normalization.}
    \label{fig:mds_icewater_wo_norm}
\end{figure}

Similarly to the results of LJ systems, when all pairwise distances shown in \Figref{fig:hist_ice_water_wo_norm} are embedded into a two-dimensional space using PCA, the result presented in \Figref{fig:mds_icewater_wo_norm} suggests that the ice and water molecule groups are clearly distinguished, and the data labeled ``ice'' are clustered into six groups corresponding to the vertices of the hexagonal structure.
This classification can be attributed to the orientation of the dipole vectors of ice (i.e., the molecular orientation of ice), which has twelve distinct directions in ice Ih \cite{Ba2011_water}.
Cross-referencing with the simulation data reveals that each cluster comprises molecules with pairs of positive and negative dipole orientations. This arises from the use of the correlation function defined in \eqref{eq:correlation_function}, in which the dipole orientations are multiplied. As a result, pairs of dipoles with opposite signs will yield the same correlation value and are therefore grouped into the same cluster.

\begin{figure}[htbp]
    \centering
    \includegraphics[width=0.75\linewidth]{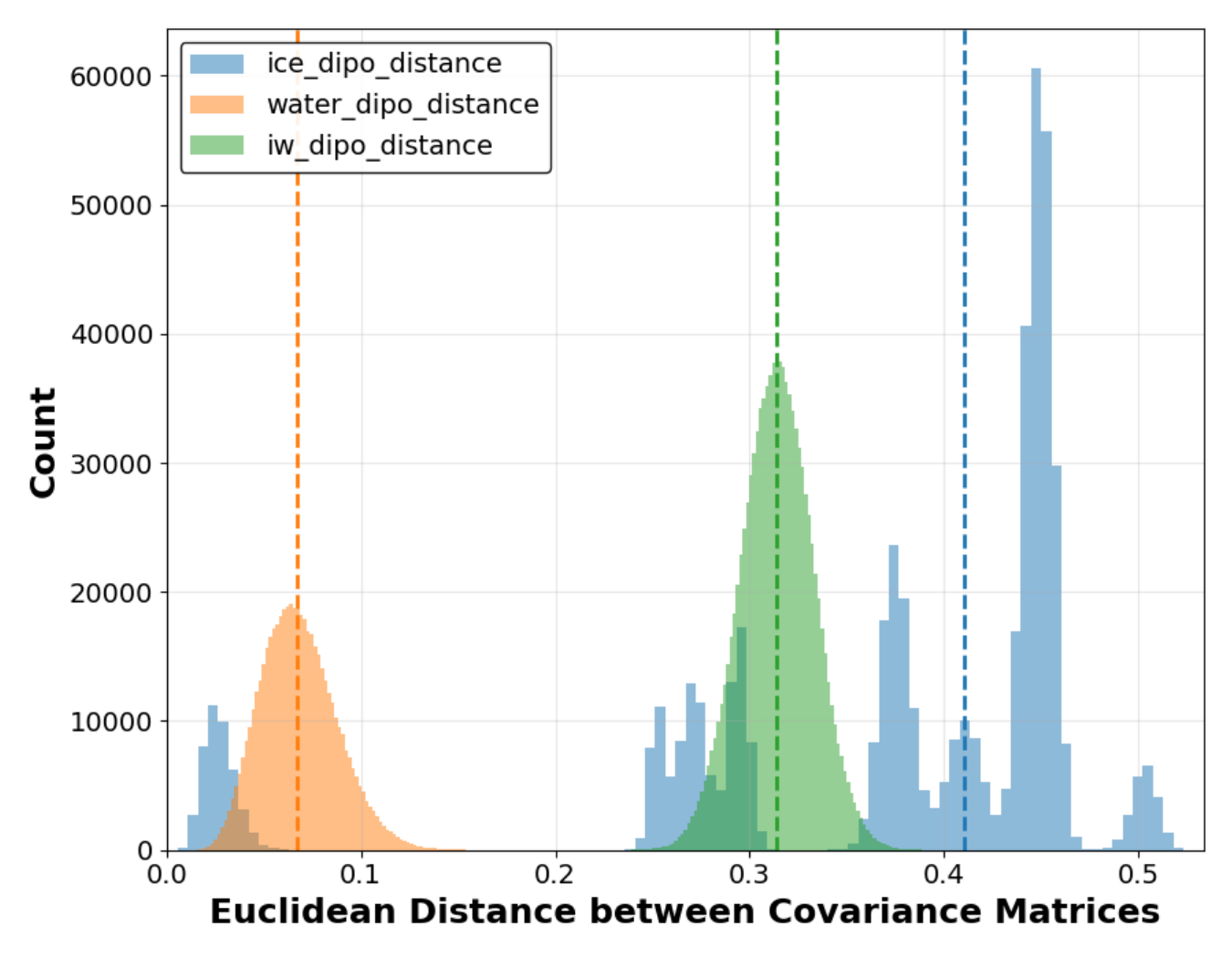}
    \caption{The three histograms show the distances between the ice system and the liquid water system: distances between ice molecules (blue), distances between liquid water molecules (orange), and distances between the two systems (green). The dotted lines in each color indicate the mean values of the corresponding histograms.}
    \label{fig:hist_ice_water_w_norm}
\end{figure}

\begin{figure}[H]
    \centering
    \includegraphics[width=0.5\linewidth]{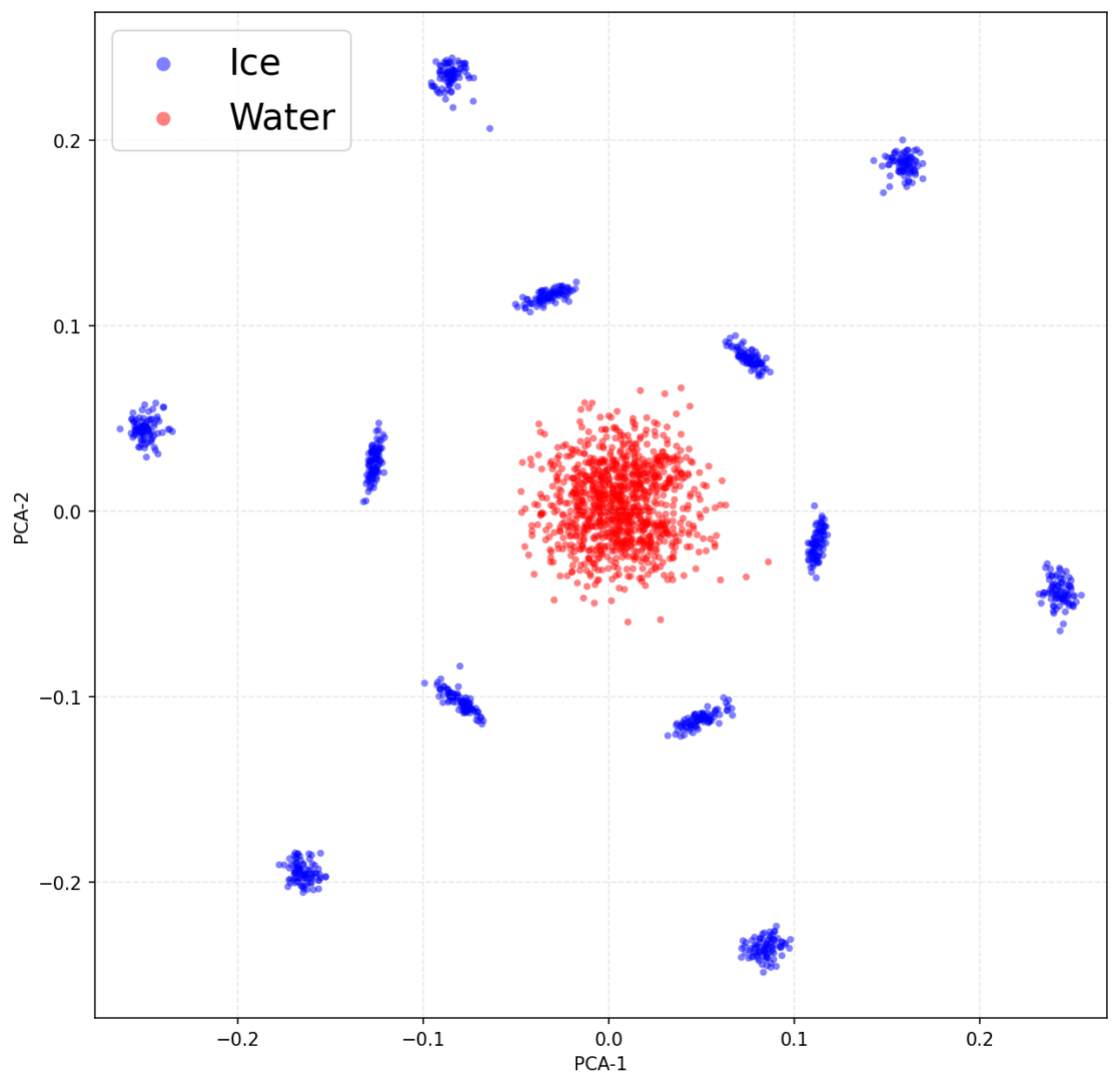}
    \caption{The two-dimensional PCA projection of all pairwise distances within and between water (red) and ice (blue) molecules with normalization for each segment of $N=8$.}
    \label{fig:mds_icewater_w_norm}
\end{figure}

\begin{figure}[H]
    \centering
    % 1枚目の画像
    \begin{subfigure}[b]{0.48\linewidth}
        \centering
        \includegraphics[width=\linewidth]{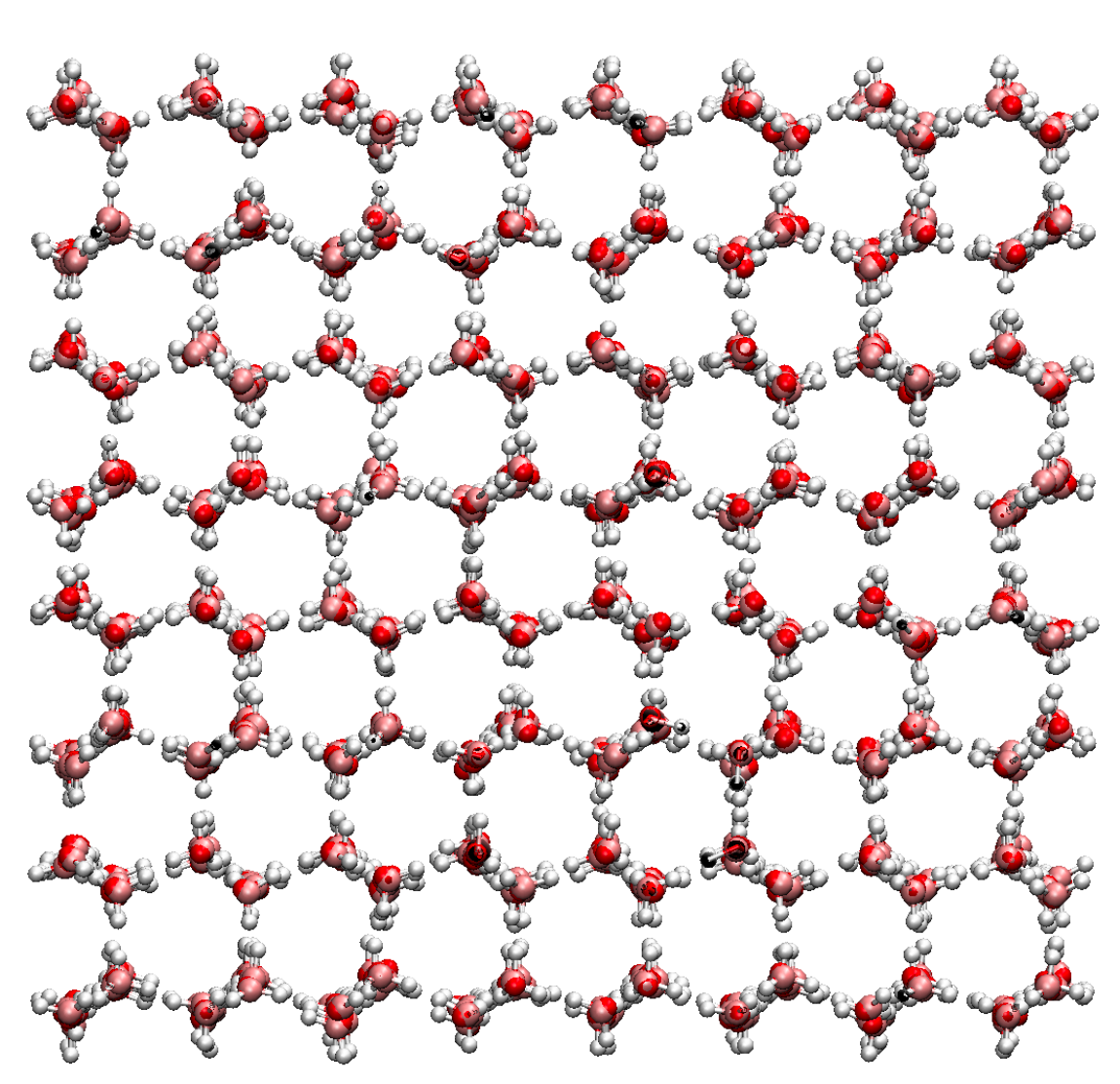}
        \caption{The $xy$-plane of the ice crystal structure}
        \label{fig:xyplane_iceih}
    \end{subfigure}
    \hfill % 2つの画像の間に適切な空白を挿入
    % 2枚目の画像
    \begin{subfigure}[b]{0.48\linewidth}
        \centering
        \includegraphics[width=\linewidth]{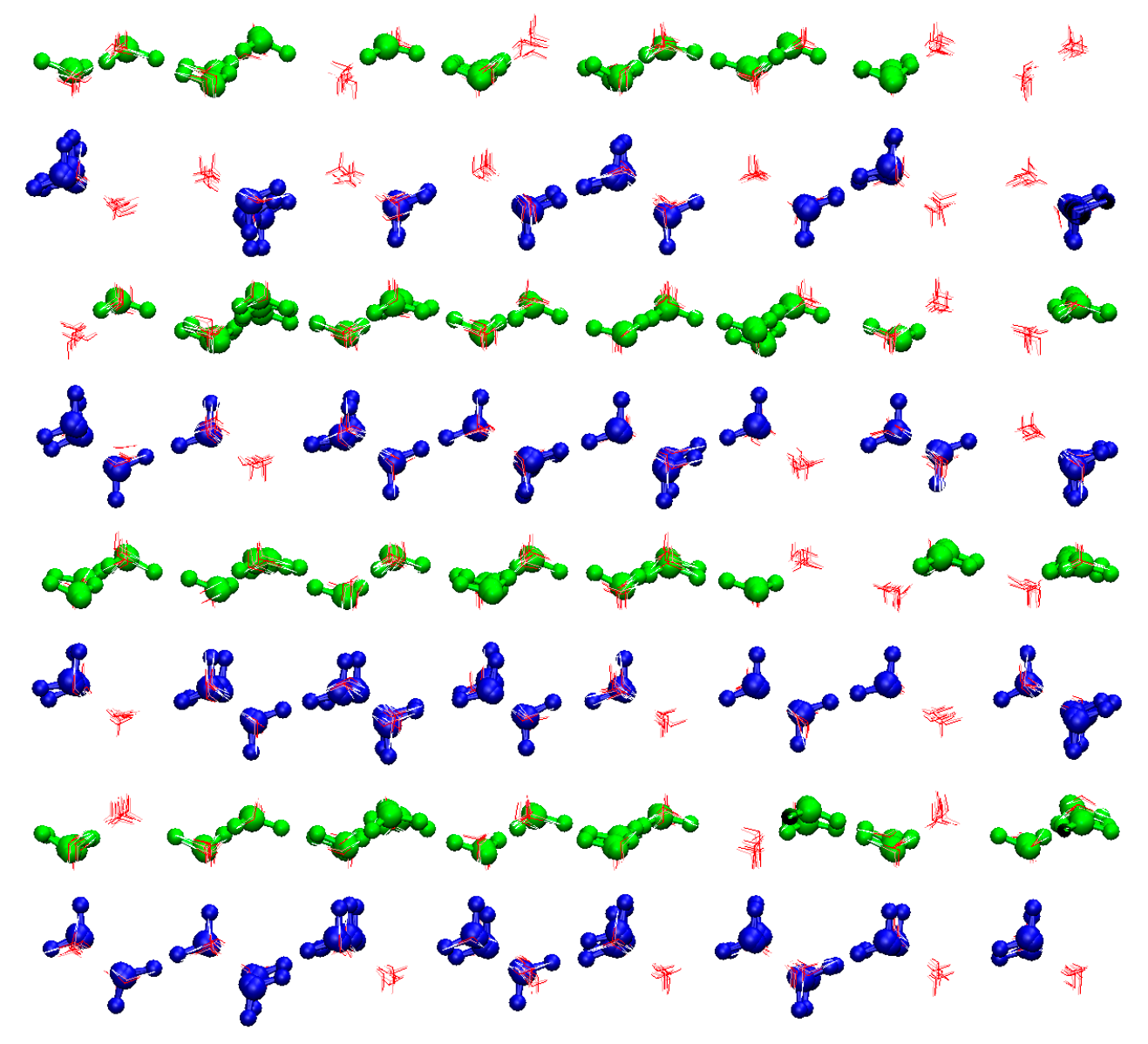} 
        \caption{Representative inner- and outer-configuration clusters mapped onto the ice crystal structure.}
        \label{fig:iceih_cluster}
    \end{subfigure}
    
    % 全体のキャプションとラベル
    \caption{Comparison between the clustering results and the ice crystal structure. The $xy$-plane of the ice crystal structure is shown in (a). In (b), representative clusters from the outer- and inner-configuration groups are mapped onto the crystal structure to visualize the difference between the two configurations.} \label{fig:ice_Ih_figure}
\end{figure}

Next, the data are normalized within each small segment of size $N=8$, and the resulting distance histogram  and two-dimensional embedding are shown in \Figref{fig:hist_ice_water_w_norm} and \Figref{fig:mds_icewater_w_norm}, respectively.
Compared with the non-normalized case, the main difference is the overall scale of the distances. In particular, the difference in the two-dimensional embedding is that the ice molecule clusters are separated into twelve types. 
These twelve types correspond to the six directions of the hexagonal structure, each of which is further divided into outer and inner hexagonal configurations. As in the previous result, the clustering of hexagonal structures is mainly governed by the dipole orientation.
To clarify the difference between the inner and outer configurations, we compare the clustering results with the crystal structure of ice Ih. The $xy$-plane of the ice crystal structure is shown in \Figref{fig:xyplane_iceih} as a representative example. Due to the symmetry of the crystal lattice, similar observations can be obtained from other crystallographic planes. When one representative group from the outer configuration group and one from the inner configuration group are mapped onto this structure, the resulting arrangement is shown in \Figref{fig:iceih_cluster}.
The six outer-configuration clusters, represented by the green cluster, contain hydrogen atoms only along the horizontal plane in the $xy$-plane. In contrast, the six inner-configuration clusters, represented by the blue cluster, contain hydrogen atoms oriented along the vertical direction. These observations indicate that subtle changes in the vibrational characteristics depend on the orientation of the bonded hydrogen atoms. By subtracting a mean, or a first moment, these differences become more distinguishable in the clustering result.


\section{Conclusions}\label{sec: conclusions}
In this study, we introduced a statistical framework for analyzing MD trajectories by quantifying the dissimilarity between system states using covariance matrices estimated from observation data. 
Numerical validation with an LJ particle system demonstrated that the proposed method effectively extracts essential dynamical features, 
revealing a strong linear correlation between the latent representation obtained from the statistical distances and the macroscopic diffusion coefficient. 
This result suggests that global transport properties are intrinsically encoded in local, short-term velocity fluctuations, enabling efficient estimation of macroscopic properties without requiring long-time trajectory integration.

The analysis of separate bulk systems of ice and liquid water further demonstrated the capability of the proposed framework to distinguish between different phases of matter. 
The results highlight the sensitivity of the covariance-based statistical distance to structural and dynamical differences in the orientation of the bonded hydrogen atoms at the molecular level.

Building on these findings, future studies will investigate phase transition phenomena in single-component molecular systems, such as the melting and freezing processes of water, 
as well as other materials exhibiting polymorphism. 
In particular, the sensitivity of the proposed framework to local dynamical fluctuations opens up possibilities for characterizing complex biomolecular processes, 
ranging from large-scale structural changes, such as protein folding, to subtle interactions, including ligand-binding events.
By capturing small variations in the covariance structure, 
the method could facilitate the identification of binding sites or the characterization of allosteric transitions, providing a computationally efficient tool for functional analysis of proteins.
Such investigations may provide deeper insights into the microscopic mechanisms governing phase transitions and support the development of predictive models for complex molecular systems.

In the present work, the Euclidean distance was employed as the statistical metric between covariance matrices because of its simplicity and computational efficiency. 
However, the proposed framework can naturally be extended by adopting metrics that respect the Riemannian manifold structure of SPD matrices. Examples include the Log-Euclidean metric, the affine invariant Riemannian metric, and the Bures--Wasserstein metric\cite{ono2024riemannian,bhatia2009positive, bhatia2019165,arsigny2007geometric}. Incorporating such geometrically consistent distances may allow the framework to capture both linear and nonlinear correlations in the underlying dynamics, providing a richer representation of molecular behavior.

Future research could further extend the methodology by incorporating higher-order statistical descriptors, such as the skewness and kurtosis of particle velocities or positions, 
to capture subtle nonlinear dynamical effects. 
In addition, although the present study focused on MD simulation data, the proposed framework is expected to be applicable to experimental datasets, 
including time-resolved spectroscopic measurements and single-molecule tracking data. 
Extending the approach to experimental systems would represent an important step toward bridging molecular simulations and real-world observations, 
potentially enabling new data-driven approaches for characterizing complex molecular dynamics.

\section*{Acknowledgements}
YO was partially supported by JST SPRING (No. Y01GQ25189) and Keio University (Doctorate Student Grant-in-Aid Program from Ushioda Memorial Fund (No. Y01JI25123)). TS was partially supported JSPS KAKENHI (No. JP25K23429). LP was partially supported by JSPS KAKENHI (No. JP24K06852), 
JST CREST (No. JPMJCR24Q5), and Keio University (Fukuzawa Fund and Academic
Development Fund).  
% This work was partially supported by JST SPRING (No. Y01GQ25189), 
% JSPS KAKENHI (No. JP24K06852 and No. JP25K23429), 
% JST CREST (No. JPMJCR24Q5), 
% and Keio University (Doctorate Student Grant-in-Aid Program from Ushioda Memorial Fund (No. Y01JI25123), Fukuzawa Fund, and Academic
% Development Fund).

% \section*{Supporting information}

% A listing of the contents of each file supplied as Supporting Information
% should be included. For instructions on what should be included in the
% Supporting Information as well as how to prepare this material for
% publications, refer to the journal's Instructions for Authors.

% The following files are available free of charge.
% \begin{itemize}
%   \item Filename-1: brief description
%   \item Filename-2: brief description
% \end{itemize}

%%%%%%%%%%%%%%%%%%%%%%%%%%%%%%%%%%%%%%%%%%%%%%%%%%%%%%%%%%%%%%%%%%%%%
%% If you are using classical BibTeX rather than biblatex,
%% remove the \printbibliography and uncomment the \bibliograpy one
%%%%%%%%%%%%%%%%%%%%%%%%%%%%%%%%%%%%%%%%%%%%%%%%%%%%%%%%%%%%%%%%%%%%%

\bibliography{citations} 
% \bibliographystyle{abbrv}
% \bibliographystyle{unsrt}
%\bibliography{MOP}
\bibliographystyle{abbrv}

\end{document}